\documentclass[12pt]{iopart}
\usepackage[T1]{fontenc}
\usepackage{graphicx}
\usepackage{color}
 \expandafter\let\csname equation*\endcsname\relax
 \expandafter\let\csname endequation*\endcsname\relax
\usepackage{amsmath}
\usepackage{amssymb}
\usepackage{bbm}

\input epsf
\begin{document}

\title[FFLO phase in superconducting nanofilms]
{Multiband effects on the Fulde-Ferrell-Larkin-Ovchinnikov state in superconducting
metallic nanofilms.}
\author{P. W\'ojcik$^1$, M. Zegrodnik$^2$}
\address{$^1$AGH University of Science and Technology, Faculty of
Physics and Applied Computer Science, al. Mickiewicza 30,
Krak\'ow, Poland}
\address{$^2$AGH University of Science and Technology,
Academic Centre for Materials and Nanotechnology, Al. Mickiewicza 30,
30-059 Krak\'{o}w, Poland}

\ead{pawel.wojcik@fis.agh.edu.pl, michal.zegrodnik@gmail.com}


\submitto{Superconductor Science and Technology}

\maketitle

\begin{abstract}
In superconducting nanofilms the energy quantization induced by the confinement in the direction
perpendicular to the film leads to a multiband character of the system which results in
the thickness-dependent oscillations of the in-plane critical field (shape resonances). 
In this paper, we study the Fulde-Ferrel-Larkin-Ovchinnikov (FFLO) phase in nanofilms and examine
its interplay with the shape resonances as well as the influence of the multiband
effects on its stability. We demonstrate that the range of the magnetic field for which the FFLO
state is stable oscillates as a function of the film thickness with the phase shift equal to one
half of the period corresponding to the critical magnetic field oscillations. Moreover, the
multiband effects lead to a division of the FFLO phase stability region into subregions number of
which depends on the number of bands participating in the formation of the paired state.
\end{abstract}


\section{Introduction}
The superconducting Fulde-Ferrel-Larkin-Ovchinnikov (FFLO) phase is characterized by the nonzero
total momentum of the Cooper pairs. According to the original
idea, proposed by Fulde and Ferrel~\cite{Fulde} as well as independently by Larkin and
Ovchinnikov~\cite{Larkin}, the nonzero momentum pairing appears in the presence of external
magnetic field as a consequence of the Fermi wave
vector mismatch between particles from two spin subbands with opposite momenta. Such unconventional
paired state can survive in magnetic fields
higher than the second critical field $H_{c2}$. In spite of many theoretical studies regarding the
FFLO state~\cite{Shimahara1994, Casalbuoni2004,Ptok2009, Maska2010}, the experimental signs of
the nonzero momentum pairing have been reported
only recently in the heavy fermion systems \cite{Matsuda2007,Blanchi2003,Kumagai2006,Correa2007}
and the two dimensional organic superconductors
\cite{Beyer2013,Singleton2000,Tanatar2002,Shinagawa2007}. The difficulties in experimental
observation of the FFLO state are caused by the stringent conditions for its appearance. First of
all, the high value of the Maki parameter is essential, which
means that the Pauli paramagnetic effect has to be strong relative to the orbital
pair-breaking mechanism being detrimental to the FFLO state~\cite{Gruenberg1966}. Moreover, the
system has to be ultra clean as the FFLO phase is easily destroyed by the
impurities~\cite{Aslamazov1969,Takada1970}.

The ultra thin metallic nanofilms which can be fabricated due to the huge development of
nanotechnology~\cite{Ozer2006,Ozer2006_b,Guo2004,Eom2006,Zhang2010,Uchihashi2011,Qin2009} may
satisfy the conditions for the FFLO phase appearance.
In the nanofilms subjected to the parallel magnetic field, the confinement in the
direction perpendicular to the film reduces the orbital effect leading to a high value of the Maki parameter.
However, if the size of the system becomes comparable to the electron's wave length, the Fermi
sphere splits into a set of discrete two-dimensional subbands leading to the multiband character of
the system. The multiband effects caused by the confinement lead to the thickness-dependent
oscillations of the superconducting properties (shape resonanses). So far, the
thickness-dependent oscillations of the energy gap~\cite{Court2007,Shanenko2007},
critical temperature~\cite{Ozer2006,Ozer2006_b,Guo2004,Eom2006,Ozer2007,Wojcik2014_1}, and critical
magnetic field~\cite{Bao2005,Sekihara2013,Wojcik2014_2, Wojcik2014_3,Shanenko2008} have been
experimentally observed, as well as theoretically studied. In spite of the mutiband behavior of a
variety of
superconductors, such as MgB$_2$ or the iron-based superconductors~\cite{Zocco2013,Terashima2013},
the FFLO phase with the inclusion of the multiband effect has been the subject of only a few papers.
Theoretical investigations regarding the FFLO phase appearance in a multiband model corresponding
to the iron-pnictides have been presented in Refs.~\cite{Ptok2013,Ptok2014}. Very recently we
have suggested that the non-zero momentum paired phase can appear in the absence of the external
magnetic field in LaFeAsO$_{1-x}$F$_x$ with a predominant interband pairing~\cite{Zegrodnik}.
Furthermore, the multiband effects on the FFLO phase in a Pauli-limiting two-band superconductor have
been theoretically studied in Refs.~\cite{Takahashi2014, Mizushima2014}. In these
studies~\cite{Takahashi2014, Mizushima2014} it has been found that a competing effect can arise from
the coupling in two bands each of which has their
own favorable momentum of the Cooper pairs. This gives rise to a rich FFLO phase diagram with the
subregions corresponding to different Cooper pair momenta. 

In the present paper we consider free-standing Pb(111) metallic nanofilms in the presence of the
in-plane magnetic field and investigate the formation of the FFLO phase as well as its interplay
with the multiband character of the system. In our considerations the number of subbands is
determined by the nanofilm thickness. We have shown that the shape resonance conditions are
detrimental to the FFLO phase formation. As a consequence, the magnetic
field for which the FFLO state is stable oscillates as a function of the film thickness with the
phase shift equal to one half of the period corresponding to the critical magnetic field
oscillations. Moreover, the multiband effects lead to a division of the FFLO phase stability region
into subregions number of which depends on the number of bands participating in the formation of the paired state.

The paper is organized as follows: in Sec.~\ref{sec:model} we introduce the basic concepts of the
theoretical scheme based on the BCS theory. In Sec.~\ref{sec:results} we present the results while 
the summary is included in Sec.~\ref{sec:concl}.


\section{Theoretical method}
\label{sec:model}
We start with the BCS Hamiltonian in the presence of external in-plane magnetic field
$\mathbf{H}_{||}=(H_{||},0,0)$ 
\begin{eqnarray}
\hat{\mathcal{H}}&=&\sum _{\sigma} \int d^3 r
\hat{\Psi}^{\dagger} (\mathbf{r},\sigma) \hat{H}_e ^{\sigma}
\hat{\Psi}(\mathbf{r},\sigma) \nonumber \\ 
&+& \int d^3 r \left [ \Delta
(\mathbf{r})\hat{\Psi}^{\dagger}(\mathbf{r},\uparrow)
\hat{\Psi}^{\dagger}(\mathbf{r},\downarrow) +H.c. \right ] 
+\int d^3r \frac{|\Delta(\mathbf{r})|^2}{g},
\label{eq:ham}
\end{eqnarray}
where $\sigma$ corresponds to the spin state $(\uparrow, \downarrow)$, $g$ is the
phonon-mediated electron-electron coupling constant, while $\Delta(\mathbf{r})$ is the
superconducting gap parameter in real space defined as
\begin{equation}
 \Delta(\mathbf{r})=-g \left < \hat{\Psi} (\mathbf{r},\downarrow)
\hat{\Psi} (\mathbf{r},\uparrow)  \right >.
\label{eq:gap_def}
\end{equation}
The single-electron Hamiltonian
$\hat{H}_e^{\sigma}$ is given by 
\begin{equation}
\hat{H}_e ^\sigma = \frac{1}{2m} \left ( -i\hbar \nabla +\frac{e}{c}
\mathbf{A} \right )^2 + s\mu_B H_{||} - \mu _F,
\label{eq:hame}
\end{equation}
where $s=+1(-1)$ for $\sigma=\uparrow (\downarrow)$, $m$ is the effective electron mass, $\mu_F$ is
the chemical potential, and we have choosen the gauge for the vector potential as
$\mathbf{A}=(0,-H_{||}z,0)$. Due to the confinement of electrons in the direction perpendicular to
the film
($z$ axis) the quantization of the energy appears. Thus, the field operators in
Eq.(\ref{eq:ham}) have the form
\begin{equation}
 \hat{\Psi}(\mathbf{r},\sigma)=\sum_{\mathbf{k}n}
\phi_{\mathbf{k}n}(\mathbf{r})\:\hat{c}_{\mathbf{k} n \sigma},\quad
\hat{\Psi}^{\dagger}(\mathbf{r},\sigma)=\sum_{\mathbf{k}n}
\phi^*_{\mathbf{k} n}(\mathbf{r})  \:
\hat{c}^{\dagger}_{\mathbf{k} n \sigma},
\label{eq:field_op}
\end{equation}
where $\hat{c}_{\mathbf{k} n \sigma}
(\hat{c}^{\dagger}_{\mathbf{k} n \sigma})$ is
the anihilation (creation) operator for an electron with spin $\sigma$ in a state characterized by
the quantum numbers $(\mathbf{k},n)$. The single-electron eigenfunctions
$\phi_{\mathbf{k}n}(\mathbf{r})$ of the Hamiltonian
$\hat{H}^{\sigma}_e$ are given below
\begin{equation}
 \phi_{\mathbf{k}n}(\mathbf{r})=\frac{1}{2\pi} e^{ik_xx}e^{ik_yy}
\varphi_{k_yn}(z),
\label{eq:ses}
\end{equation}
where $\mathbf{k}=(k_x,k_y)$ is the electron wave vector, while $n$ labels the discrete quantum
states. We determine $\varphi_{k_yn}(z)$ by diagonalization of the Hamiltonian (\ref{eq:hame}) in
the basis of the quantum well states 
\begin{equation}
 \varphi_{k_yn}(z)= \sqrt{\frac{2}{d}} \sum _{l} c_{k_yl} \sin \left [
\frac{\pi(l+1)z}{d} \right ],
\end{equation}
where $d$ is the nanofilm thickness. We use the hard-wall potential as the boundary condition for the wave function in the $z$-direction.

In the FFLO phase, induced by the magnetic field, the Cooper pairs gain the nonzero total
momentum $\mathbf{q}$ which results from the pairing  of electrons from spin-splitted subbands
$[(\mathbf{k},n),(-\mathbf{k}+\mathbf{q}n)]$.
For such pairing, the substitution of the expression for the gap parameter \ref{eq:gap_def}) and the
field operators (\ref{eq:field_op}) into Eq. (\ref{eq:ham}) gives the following form of the
Hamiltonian
\begin{eqnarray}
 \hat{\mathcal{H}}&=& \sum _{\mathbf{k}n}  
 \begin{array}{c}
 ( \hat{c}^{\dagger} _{\mathbf{k}n\uparrow} \: \hat{c}
_{-\mathbf{k}+\mathbf{q} n \downarrow} ) \\
 \\
 \end{array}
 \left (
 \begin{array}{cc}
 \xi _{\mathbf{k} n} &  \Delta _{\mathbf{q}n} \\
 \Delta _{\mathbf{q}n} & -\xi _{-\mathbf{k}+\mathbf{q} n} 
 \end{array}
 \right )
 \left (
 \begin{array}{c}
  \hat{c} _{\mathbf{k} n \uparrow}  \\
  \hat{c} ^ { \dagger} _{-\mathbf{k}+\mathbf{q}n \downarrow} 
 \end{array} 
 \right ) \nonumber \\
 &+& \sum _{\mathbf{k}n} \xi _{-\mathbf{k}+\mathbf{q} n} +\sum _{n}
\frac{|\Delta _{\mathbf{q}n}|^2}{g},
\label{eq:ham_2}
\end{eqnarray}
where $\mathbf{q}$ is the total momentum of the Cooper pairs in the $(x,y)$ plane. Analogously as
in the paper by Fulde and Ferrell \cite{Fulde} we have assumed that all the Cooper pairs have the
same momentum $\mathbf{q}$. The direction of $\mathbf{q}$ is arbitrary for the case of s-wave
pairing symmetry with parabolic dispersion relation. For simplicity we take on $\mathbf{q}=(q,0)$.
The energy gap in reciprocal space is defined as
\begin{equation}
\Delta_{\mathbf{q}n'}=\frac{g}{4 \pi ^2} \sum _{\mathbf{k}n}
C_{\mathbf{k} n'n} \langle\hat{c}_{-\mathbf{k}+\mathbf{q}n\downarrow}\hat{c}_{\mathbf{k}n\uparrow}
\rangle,
\end{equation}
where 
\begin{equation}
 C_{\mathbf{k}n'n}=\int dz \varphi_{k_y n'}(z) \varphi_{-k_y n'}(z)
\varphi_{k_y n}(z) \varphi_{-k_y n}(z).
\label{eq_inter}
\end{equation}
The Hamiltonian (\ref{eq:ham_2}) can be reduced to the diagonal form by the
Bogoliubov-de Gennes transformation $\hat{c}_{\mathbf{k} n \sigma} =
u_{kn\sigma}\gamma_{kn}+s v^*_{kn\sigma}\gamma^{\dagger}_{kn}$~\cite{Gennes}. As a
result, one obtains the following form of the
quasiparticle energies
\begin{equation}
 E^{\pm}_{\mathbf{k}\mathbf{q} n}=\frac{1}{2} \left ( \xi _{\mathbf{k} n}-\xi
_{-\mathbf{k}+\mathbf{q} n} \pm \sqrt{\alpha_{\mathbf{k}\mathbf{q} n }} \right ) +
 \mu_B H,
\end{equation}
 with 
\begin{equation}
 \alpha_{\mathbf{k}\mathbf{q}n} = \left ( \xi _{\mathbf{k} n}+\xi _{-\mathbf{k}+\mathbf{q} n}
\right ) ^2 +4\Delta_{\mathbf{q}n} ^2.
\end{equation}
The self-consistent equation for the superconducting gaps (for $n$ bands we have $n$ gap
parameters) has the form
\begin{eqnarray}
\Delta_{\mathbf{q}n'}&=&\frac{g}{4 \pi ^2} \sum _{\mathbf{k}n}
C_{\mathbf{k} n'n} \frac{\Delta_{\mathbf{q}n}}{\sqrt{\alpha_{\mathbf{k}\mathbf{q}n}}}
\left [ 1-f(E^{+}_{\mathbf{k}\mathbf{q} n}) - f(E^{-}_{\mathbf{k}\mathbf{q} n}) \right ],
\label{eq:delta_fflo}
\end{eqnarray}
where $f(E)$ is the Fermi-Dirac distribution. The summation in
Eq.~(\ref{eq:delta_fflo}) is carried out only
over the single-electron states with energy $\xi _{\mathbf{k} n}$
inside the Debye window $\left | \xi _{\mathbf{k} n}  \right | <
\hbar \omega _D$, where $\omega _D$ is the Debye frequency.
Since the chemical potential $\mu$ for nanofilms can strongly deviates from its bulk value
$\mu_{bulk}$ for each nanofilm thickness we calculate $\mu$ using the expression
\begin{eqnarray}
n_e&=&\frac{1}{d} \sum _{\mathbf{k}n}\int _0^d dz
\bigg \{ |u_{\mathbf{k}\mathbf{q}n}\varphi_{k_yn}(z) |^2f(E_{\mathbf{k}\mathbf{q}n}) \nonumber \\
&+&|v_{\mathbf{k}\mathbf{q}n} \varphi_{k_yn}(z)|^2 [1-f(E_{\mathbf{k}\mathbf{q}n})]\bigg \},
\label{eq:mu}
\end{eqnarray}
where $n_e$ is the electron density corresponding to the bulk value (it refers to the chemical
potential $\mu_{bulk}$) and $u_{\mathbf{k}\mathbf{q}n}$,
$v_{\mathbf{k}\mathbf{q}n}$ are the Bogolubov coherence factors.

After calculating the superconducting gaps in the reciprocal space $\Delta_{\mathbf{q}n}$ one can
determine the spatial dependence of the order parameter by using the following formula
\begin{equation}
 \Delta_{\mathbf{q}}(\mathbf{r})=e^{i\mathbf{q}\cdot\mathbf{r}}\Delta_{\mathbf{q}}(z),
\end{equation}
where
\begin{equation}
\Delta_{\mathbf{q}}(z)=\frac{g}{4 \pi^2} \sum_{\mathbf{k}n} \varphi_{k_y n}(z) \varphi_{-k_y n}(z) \frac{\Delta_{\mathbf{q}n}}{\sqrt{\alpha_{\mathbf{k}\mathbf{q}n}}}
\left [ 1-f(E^{+}_{\mathbf{k}\mathbf{q} n}) - f(E^{-}_{\mathbf{k}\mathbf{q} n}) \right ].
\label{delta}
\end{equation}
As one can see the spatial dependence of the superconducting gap in the $(x,y)$ plane is due to the nonzero center of mass momentum of the Cooper pairs while the $z$-dependence comes from the confinement of the electrons within the nanofilm.
In the further analysis we often use the spatially averaged energy gap
defined as
\begin{equation}
 \bar{\Delta}_{\mathbf{q}}=\frac{1}{d} \int _0 ^d \Delta_{\mathbf{q}}(z) dz.
\end{equation}
To obtain the results presented in the next section, the set of self-consistent equations
(\ref{eq:delta_fflo}), (\ref{eq:mu}) is calculated numerically and the Cooper pair total momentum
$\mathbf{q}$ is determined by minimizing the free energy of the system~\cite{Kosztin}.


\section{Results}
\label{sec:results}
In the present paper we consider the FFLO phase induced by the in-plane magnetic field in
free-standing Pb(111) nanofilms.
The first-principle calculations of the electronic structure 
for Pb nanofilms~\cite{Wei2002,Wei2007,Miller2009} demonstrated that the quantum size effect
for Pb(111) can be well described by the quantum well states centered at the L-point of a
two-dimensional Brillouin zone~\cite{Wei2002}. Moreover, the energy dispersion calculated for
Pb(111) nanofilms is nearly parabolic~\cite{Wei2002}. Based on these results, in our calculations 
we use the parabolic band approximation treating the bulk Fermi level $\mu _{bulk}$ and the
electron mass $m$ as the fitting parameters. Their values are determined based on the results from
the first principle calculations presented in Refs.~\cite{Wei2002}. In our study we use the
following values of the parameters: $gN_{bulk}(0)=0.39$ where
$N_{bulk}(0)=mk_F/(2 \pi^2 \hbar ^2)$  is the bulk density of the single-electron states at the
Fermi level, $\hbar \omega _D=8.27$~meV, the bulk critical temperature $T_{bulk}=7.2$~K
corresponding to the energy gap $\Delta_{bulk}=1.1$~meV and $\mu _{bulk}=3.8$~eV which corresponds
to the electron density $n_e=4.2 \times 10^{21}$~cm$^{-3}$. Since the calculations for the FFLO
phase in nanofilms are time-consuming we restrict our analysis to the temperature $T=0$~K.  
\begin{figure}[ht]
\begin{center}
\includegraphics[scale=0.4]{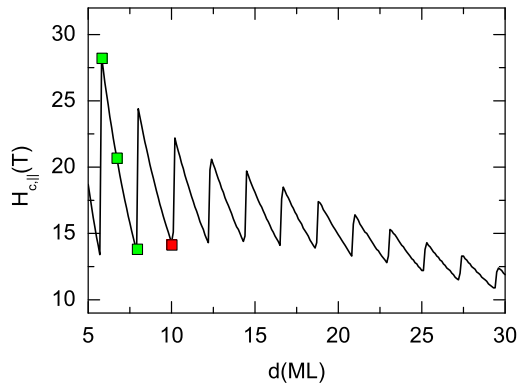}
\caption{(Color online) In-plane critical magnetic field $H_{c,||}$ as a
function of the nanofilm thickness $d$ calculated without the inclusion of the FFLO phase for 
$\mathbf{q}=0$ and $T=0$~K. Squares denote the film thicknesses chosen for the analysis of the
FFLO phase stability.
}
\label{fig1}
\end{center}
\end{figure}

Figure ~\ref{fig1} displays the in-plane critical field $H_{c,||}$ as a function of the nanofilm
thickness $d$ calculated assuming $\mathbf{q}=0$ (no FFLO phase).
The 'tooth-like' oscillations presented in Fig.~\ref{fig1} are well
known~\cite{Wojcik2014_1,Shanenko2008} and can be explained on the basis of the electron energy
quantization. If the confinement of the electron motion in the direction perpendicular to the film
becomes comparable to the electrons wave length, the Fermi sphere splits into a set of discrete
two-dimensional parabolic subbands. The energies of these subbands decrease with increasing nanofilm
thickness. Each time a subband passes through the Fermi level the density of states in the
energy window $\left [ \mu - \hbar \omega_D,  \mu + \hbar \omega _D \right ]$, in which the
phonon-mediated pairing occurs, abruptly increases leading to the increase of the critical magnetic
field. This phenomenon is called 'shape resonance'. In the further analysis we use the term
resonant and non-resonant thickness to define the thickness for which the resonance conditions are
satisfied or not satisfied, respectively. Therefore, the
thickness-dependent oscillations of $H_{c,||}$ presented in Fig.~\ref{fig1} are caused by the
subsequent subbands passing through the Fermi level while the nanofilm thickness is increased. A
detailed analysis of this effect can be found in our recent paper~\cite{Wojcik2014_3}.

\begin{figure}[ht]
\begin{center}
\includegraphics[scale=0.5]{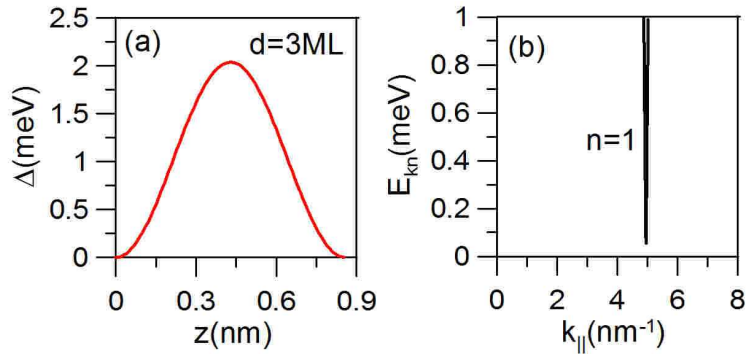}
\caption{(Color online) (a) Spatially dependent superconducting gap
$\Delta (z)$ and (b) quasi-particle band which participates in the formation of the superconducting
state. Results for the nanofilm thickness $d=3$~ML and $H_{||}=0$.
}
\label{fig2}
\end{center}
\end{figure}

\begin{figure}[ht]
\begin{center}
\includegraphics[scale=0.5]{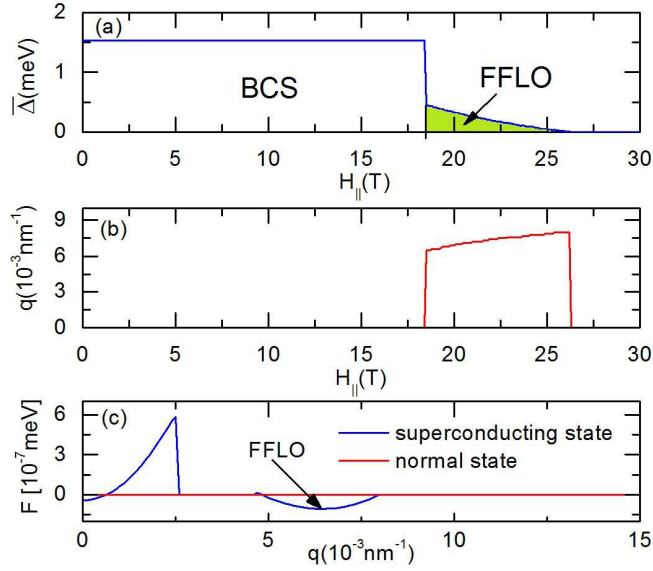}
\caption{(Color online) (a) Spatially averaged energy gap $\bar{\Delta}$
as a function of the magnetic field $H_{||}$ (the FFLO phase stability
is marked by the green region). (b) Total momentum of the Cooper pairs
$q$, which corresponds to the stable FFLO state, as a function of the magnetic
field $H_{||}$. (c) Free energy $F$ of the superconducting and
normal states as functions of the Cooper pairs total momentum $q$.
Results for the nanofilm thickness $d=3$~ML.
}
\label{fig3}
\end{center}
\end{figure}

First, we discuss the ultrathin nanofilm with the thickness $d=3$~ML (not marked in
Fig.~\ref{fig1}). For such a thin film the in-plane magnetic field acts only on the spins of the
conduction electrons and all orbital effects can be neglected. The corresponding quasiparticle
dispersion relation and the $z$-dependence of the superconducting gap are shown in Fig.~\ref{fig2}.
For the chosen thickness only the lowest band participates in the paired phase, therefore, the
multiband effects on the superconducting state do not appear.
In Fig.~\ref{fig3}(a) we plot the spatially averaged energy
gap $\bar{\Delta}$ as a function of the magnetic field $H_{||}$. The values of the Cooper pair
momentum $q$ which correspond to the stability of the FFLO phase are shown in Fig.~\ref{fig3}(b) and
have been determined by minimizing the free energy $F$ of the system.
 The dependences $F(q)$ for
the superconducting and normal phases calculated for $H_{||}=18.5$~T are
displayed in Fig.~\ref{fig3}(c). The free energy minimum, marked by arrow, corresponds to stable FFLO state. Fig.~\ref{fig3}(a)
shows that the transition form the BCS to the FFLO phase has a discontinuous
nature whereas in the FFLO region, the spatially averaged energy gap
decreases almost linearly with increasing magnetic field. The presented
behavior agrees with the one reported for single-band $s$-wave
superconductors~\cite{Matsuda2007}.

\begin{figure}[ht]
\begin{center}
\includegraphics[scale=0.5]{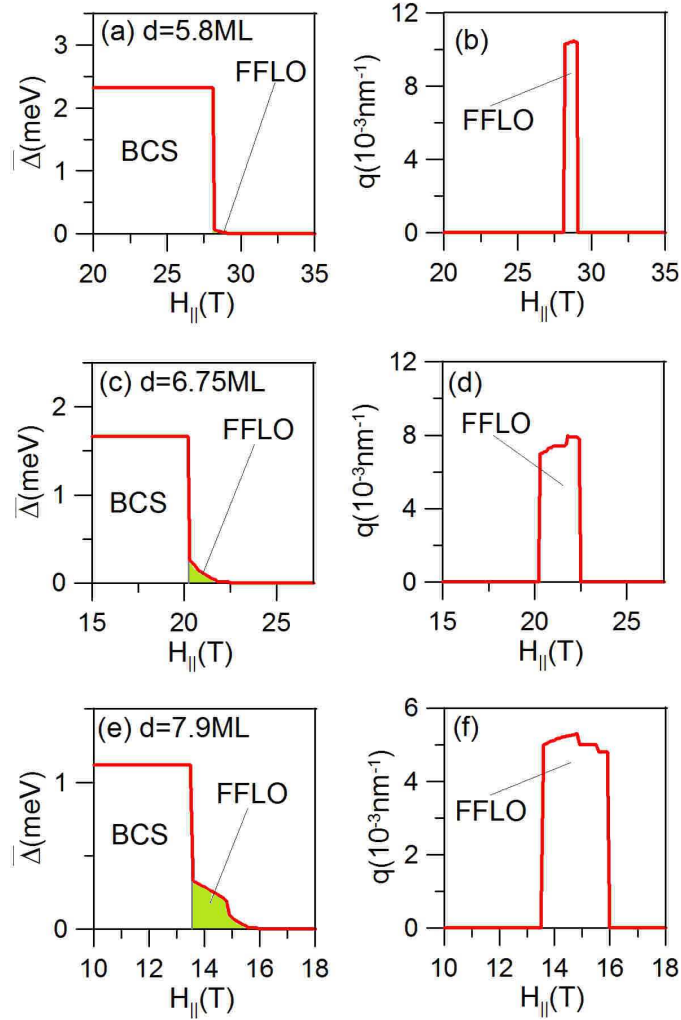}
\caption{(Color online) (a) Spatially averaged energy gap $\bar{\Delta}$
(left panels) and total momentum of Cooper pairs (right panels) as
functions of the magnetic field $H_{||}$ for the nanofilm thickness
(a,b) $d=5.8$~ML, (c,d) $d=6.75$~ML and (e,f) $d=7.9$~ML. The
chosen nanofilm thicknesses are marked by the green squares in
Fig.~\ref{fig1}.
}
\label{fig4}
\end{center}
\end{figure}

As explained above, in nanofilms, the number of bands participating in the superconducting state depends on the thickness, $d$.
If we increase the nanofilm thickness, the
subsequent subbands begin to participate in the superconducting phase
giving rise to the shape resonances presented in Fig.~\ref{fig1}. 
In order to analyze the interplay between the shape resonances and the
stability of the FFLO phase, in Fig.~\ref{fig4} (left panels) we present the
spatially averaged energy gap $\bar{\Delta}$ as a function of the
magnetic field $H_{||}$ for different nanofilm
thicknesses: (a) $d=5.8$~ML which
corresponds to the shape resonance, (b) $d=6.75$~ML in the middle
between two adjacent resonances and (c) $d=7.9$~ML for which the
resonance conditions are not satisfied. The nanofilm thicknesses chosen
for the analysis are marked by the green squares in Fig.~\ref{fig1}.
Right panels in Fig.~\ref{fig4} present the total
momentum of the Cooper pairs which correspond to the stability of the FFLO phase.
In comparison with the results for the thickness $d=3$~ML (see
Fig.~\ref{fig3}) the stability range of the FFLO phase is narrowed down due to the
multiband and orbital effects. From the presented results one can see
that the range of the FFLO phase stability is minimal for the thickness
corresponding to the shape resonance $d=5.8$~ML [see
Fig.~\ref{fig3}(a)], then it gradually increases to its maximal value
for $d=7.9$~ML, for which the shape resonance conditions are not
satisfied. The same behavior is observed for other resonances depicted
in Fig.~\ref{fig1}. Based on these results one can distinguish two characteristic
features: (i) the shape resonance conditions are detrimental to the FFLO state formation, and (ii)
the range of the magnetic field for which the FFLO phase is stable
oscillates as a function of the film thickness with the phase shift equal to one half of the period
corresponding to the critical field oscillations. 

The presented behavior can be explained on the basis of the multiband nature of the considered system.
\begin{figure}[ht]
\begin{center}
\includegraphics[scale=0.5]{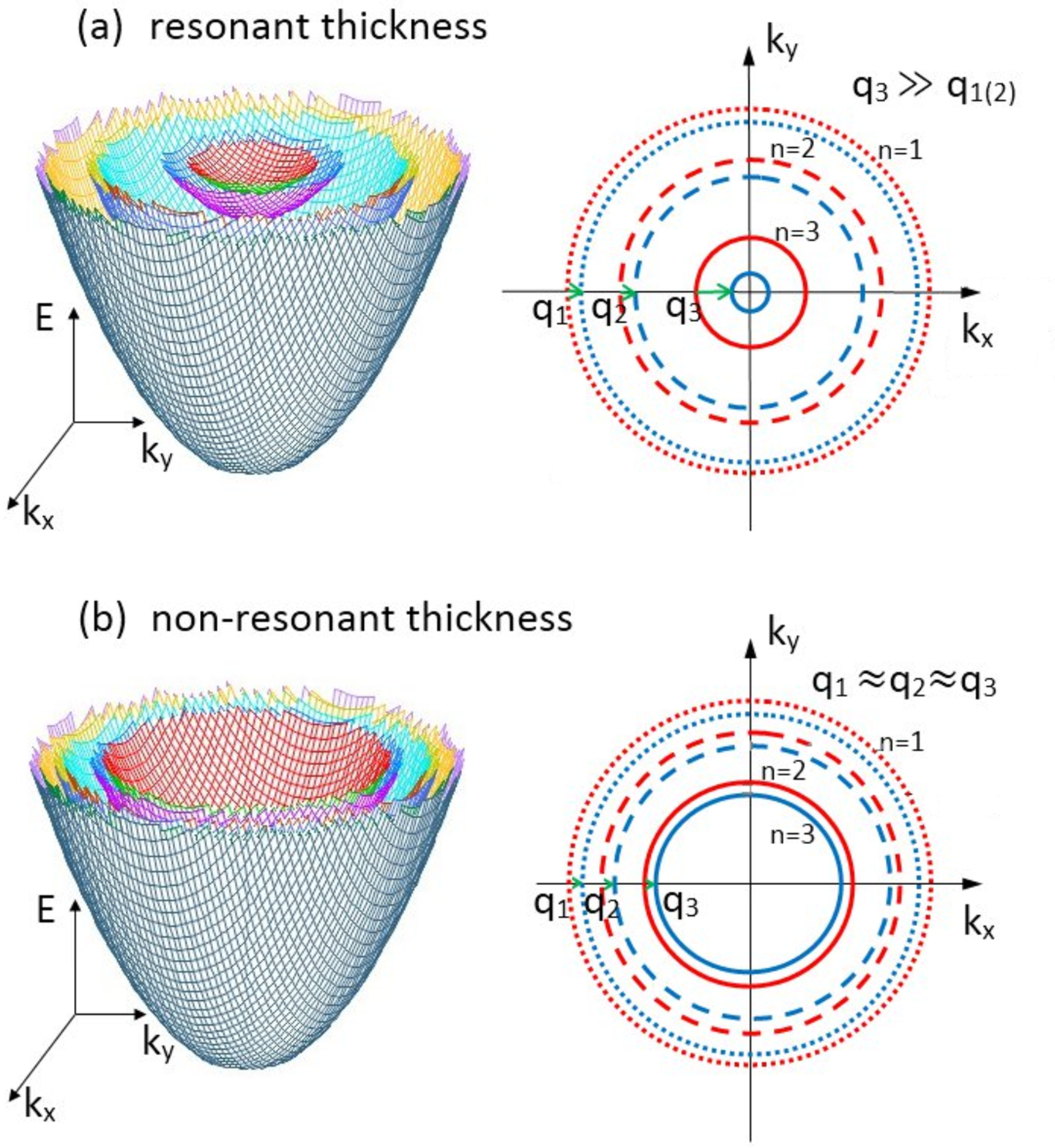}
\caption{(Color online) Schematic illustration of the single electron 
parabolic subbands (left panels) and  their intersections at the Fermi
energy (right panels) for (a) resonant thickness and (b) non-resonant
thickness. The red and blue color in the right panel corresponds to the spin-up
and spin-down subbands, respectively. The Fermi wave vector
mismatch between spin-splitted subbands, induced by the magnetic field,  can be
compensated by the non-zero total momentum of the Cooper pairs. The
Cooper pair momenta $q_n$ corresponding to all subbands are marked by green arrows in the right panels.
}
\label{fig5}
\end{center}
\end{figure}
For a given nanofilm thickness a particular number of subbands
participate in the superconducting state. In the presence of the magnetic field the wave vector
mismatch (induced by the Zeeman spin splitting), which corresponds to different subbands, may vary
significantly. As a consequence, each subband
has its own favorable value of the Cooper pair total momentum $q_n$. However, the situation in which
there is a number of independent modulation vectors $q_n$ simultaneously, each corresponding to a
different subband, cannot be realized in the system because all the subband gap parameters
$\Delta_{\mathbf{q}n}$ are coupled (c.f. Eq. \ref{eq:delta_fflo}). Since
there can be only one value of the Cooper pair total momentum in the system, a
competition between the subbands appears. For the resonant thicknesses there is one leading quasi-particle
branch $n'$ (with the energy gap $\Delta _n'$) which is responsible for
the enhancement of the paired phase~\cite{Shanenko2007}. Its
dominant influence on the formation of the superconducting state results form the high density of
states at the Fermi level which appears each time when the bottom of a
subband is close to the Fermi level. However, the closer to the Fermi
level the bottom of a subband is, the grater is the wave vector mismatch which corresponds to this
subband in the presence of magnetic field. Therefore, for the
leading quasi-particle branch the favored Cooper pair momentum $q_{n'}$ is significantly greater
than the ones which correspond to the remaining bands, as schematically
presented in Fig.~\ref{fig5}(a). Due to this fact it is not possible to adjust a single Cooper pair
total momentum in order to compensate all the wave vector mismatches, what leads to the suppression
of the FFLO phase. On the other hand, for the film thicknesses for which the resonance
conditions are not fulfilled (non-resonant thickness), the
contribution to the superconducting BCS phase coming form all the subbands
is comparable. Since the energies corresponding to the bottoms of all the subbands are far below the
Fermi level the total momenta of the Cooper pairs $q_n$ which compensate the wave vector mismatches
for all the subbands differ only slightly as presented in Fig.~\ref{fig5}(b). In such situation it
is possible to adjust the $q$ vector to compensate each Fermi wave vector mismatch to a large
extent. In this case, the FFLO phase stability range reaches its maximal value.

\begin{figure}[ht]
\begin{center}
\includegraphics[scale=0.5]{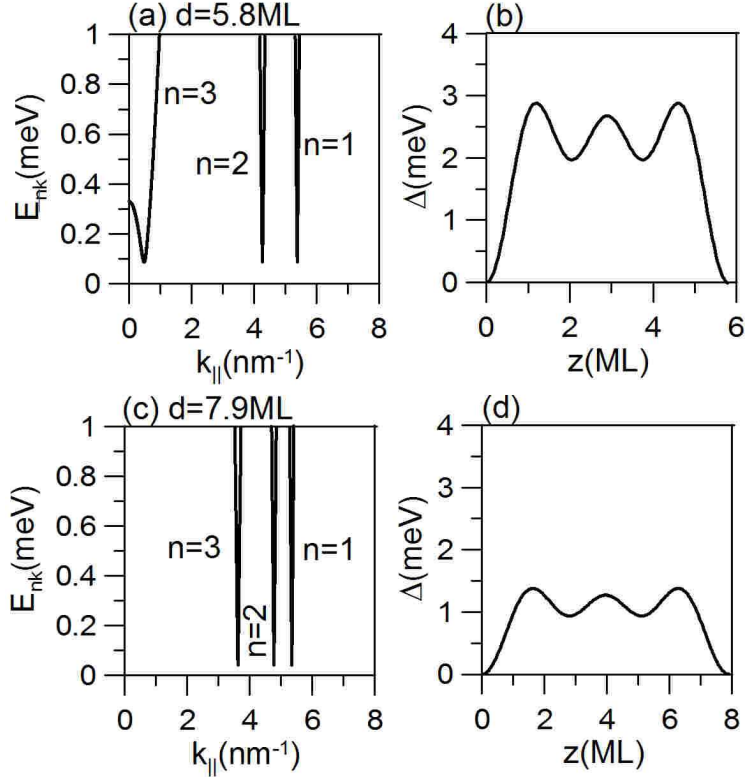}
\caption{(Color online) Quasi-particle subbands $E_{kn}$ and spatially
dependent energy gap for (a) the resonant thickness $d=5.8$~ML and (b)
the non-resonant thickness $d=7.9$~ML.
}
\label{fig6}
\end{center}
\end{figure}

In Fig.~\ref{fig6} we present the quasi-particle branches and the spatially
dependent energy gap for two thicknesses corresponding to Fig.~\ref{fig4}, $d=5.8$~ML (resonant
thickness) and
$d=7.9$~ML (non-resonant thickness) calculated for 
$H_{||}=0$ (due to the symmetry $k_x-k_y$ for $H_{||}=0$ we can define
$k^2_{||}=k_x^2+k_y^2$). 
As we can see the three lowest bands contribute to
the superconducting state for both thicknesses. The shape resonance for
$d=5.8$~ML results from the pairing in the subband $n=3$, as the bottom
of this subband has just passed through the Fermi level. As
aforementioned, this dominant resonant subband leads to the enhancement
of the energy gap [compare Fig.\ref{fig6}(b) and (d)] and is detrimental to the
FFLO state formation. On the other hand, for the non-resonant thickness
$d=7.9$~ML the three subbands are far below the Fermi energy. 
Since they have comparable total momentum of the Cooper pairs the
adjustment of a single $q$ vector for all the bands is possible giving raise to the
FFLO state stability region.

\begin{figure}[ht]
\begin{center}
\includegraphics[scale=0.5]{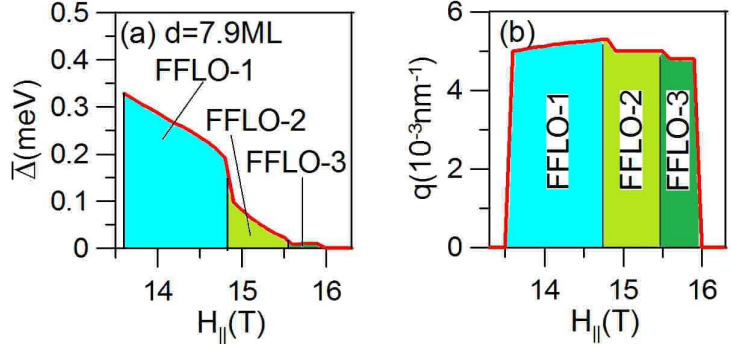}
\caption{(Color online) (a) Spatially averaged energy gap $\bar{\Delta}$
and (b) total momentum of the Cooper pairs $q$ corresponding to stable
FFLO state as a function of the magnetic field $H_{||}$. Results
presented for the same data as in Fig.~\ref{fig4}(e) but limited to 
the FFLO phase stability region. Different FFLO subphases are marked by different colors and
labeled by FFLO-$n$ ($n=1,2,3$).
}
\label{fig7}
\end{center}
\end{figure}

\begin{figure}[ht]
\begin{center}
\includegraphics[scale=0.6]{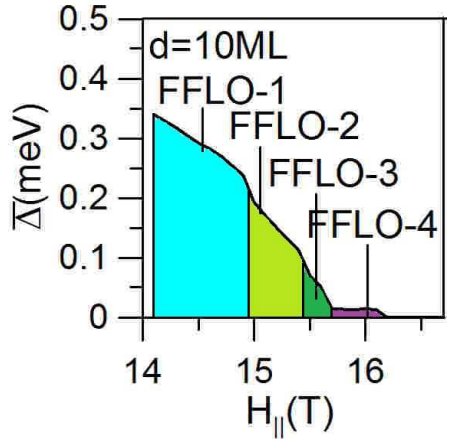}
\caption{(Color online) (a) Spatially averaged energy gap $\bar{\Delta}$ as a function of the
magnetic field $H_{||}$ for the thickness $d=10$~ML (marked by red square in Fig.~\ref{fig1}).
Results limited to the FFLO phase stability region. Different FFLO subphases are marked
by different colors and labeled by FFLO-$n$ ($n=1,2,3,4$).
}
\label{fig8}
\end{center}
\end{figure}

As it has already been mentioned for the case of multiband superconductors the competing effect leads to a complex structure of the FFLO state stability region, which is
divided into $N$ subregions, with $N$ being the number of superconducting subbands. Such behavior is
presented in Fig.~\ref{fig7} which
shows the data from Fig.\ref{fig4}(e) but limited to the FFLO phase stability region.
In Fig.~\ref{fig7} one can distinguish between three FFLO phases, labeled by
FFLO-$n$ (n=1,2,3). For a given FFLO-$n$ phase the total momentum $q$ of the Cooper
pairs is mainly determined by the Fermi wave vector mismatch in the $n$-th band. Similar behavior is observed for the
rest of non-resonant thicknesses. In Fig.~\ref{fig8} we present the FFLO
stability region for $d=10$~ML which corresponds to the non-resonant
thickness in which four subbads participate in the superconducting
state (denoted by the red square in Fig.~\ref{fig1}). By
analogy, the FFLO stability region is now divided into four subregions.
 
Further increase of the thickness makes the orbital effect more relevant what is detrimental to the
FFLO state formation. Due to the orbital effect the range of the magnetic field in which the
FFLO state exists gradually decreases with increasing film thickness and for $d$ greater than
$15$~ML this phase completely disappears.


\section{Summary}
\label{sec:concl}
The multiband effects on the FFLO state induced by the in-plane magnetic field in
Pb(111) nanofilms has been investigated in the framework of the BCS theory. 
We have studied the interplay between the FFLO phase and the shape resonances, as well as the
influence of the multiband and orbital effects on its stability. We have found that for the
resonant thicknesses the FFLO phase stability region is suppressed due the difference between the
Fermi wave vector mismatch corresponding the resonant subband and the one corresponding to the
remaining subbands. As a consequence, the range of the magnetic field for which the FFLO phase is
stable oscillates as a function of the film thickness with the phase shift equal to one half of the
period corresponding to the critical field oscillations. Moreover, we have found that
the FFLO stability region is divided into $N$ subregions where $N$ is the number of subbands which
contribute to the superconducting state.

It is worth mentioning that so far there have been no reports on the FFLO phase stability in
metallic nanofilms. This fact is mainly due to the quality of the films most of which
are in the dirty limit while the non-zero momentum paired phase is easily destroyed by the
nonmagnetic impurities~\cite{Aslamazov1969,Takada1970}. Other experimental difficulties such as
surface roughness or the magnetic field direction, which has to be applied precisely parallel to
the plane, are also detrimental to the FFLO phase formation. Nevertheless, one can hope that the
technological progress will allow for the fabrication of very clean nanofilms in the future and
investigate experimentally some of the effects studied theoretically in this paper.


\section*{Acknowledgments}
This work  was financed from the budget for Polish Science in the years 2013-2015. Project number:
IP2012 048572. M. Z. acknowledges the financial support from the Foundation for Polish Science (FNP)
within project TEAM.

\section*{References}

\end{document}